\documentclass[a4paper]{article}

\usepackage{graphics}
\usepackage{graphicx}
\usepackage{amssymb}
\usepackage{authblk} 
\usepackage{fullpage}

\begin{document}

\title{Nanocrystalline Zr$_{3}$Al Made through Amorphization by Repeated Cold Rolling and Followed by Crystallization}

\author[1]{D. Geist\footnote{corresponding author: david.geist@univie.ac.at}}
\author[2]{S. Ii}
\author[2]{K. Tsuchiya}
\author[1]{H.P. Karnthaler}
\author[3]{G. Stefanov}
\author[1]{C. Rentenberger}
\affil[1]{Physics of Nanostructured Materials, Faculty of Physics, University of Vienna, 1090 Vienna, Austria}
\affil[2]{Hybrid Materials Center, National Institute for Materials Science, Tsukuba 305-0047, Japan}
\affil[3]{Institute of Metal Science Acad. Angel Balevski, Bulgarian Academy of Sciences, 1574 Sofia, Bulgaria}

\maketitle

\begin{abstract}
The intermetallic compound Zr$_{3}$Al is severely deformed by the method of repeated cold rolling. By X-ray diffraction it is shown that this leads to amorphization. TEM investigations reveal that a homogeneously distributed debris of very small nanocrystals is present in the amorphous matrix that is not resolved by X-ray diffraction. After heating to 773 K, the crystallization of the amorphous structure leads to a fully nanocrystalline structure of small grains (10 - 20 nm in diameter) of the non-equilibrium Zr$_{2}$Al phase. It is concluded that the debris retained in the amorphous phase acts as nuclei. After heating to 973 K the grains grow to about 100 nm in diameter and the compound Zr$_{3}$Al starts to form, that is corresponding to the alloy composition.

\end{abstract}

\twocolumn

\section{Introduction}
\label{intro}

The intermetallic compound Zr$_{3}$Al has been of special interest as a nuclear structural material although a drawback of the material is embrittlement under certain irradiation conditions \cite{Schulson1995}. Recently, it was shown that nanostructuring of materials can enhance the irradiation resistance, e.g. for the intermetallic compound NiTi \cite{Kilmametov2008}. One successful method to achieve nanostructuring in bulk materials is severe plastic deformation (SPD). The formation of the nanocrystalline structure can occur directly by grain refinement of the coarse grained material or by crystallization of SPD induced amorphous material. For example, crystallization of the intermetallic alloy NiTi amorphized by severe plastic deformation can lead to nanocrystalline structures and by modification of the deformation path and the heat treatment, properties of the alloy can be tailored \cite{Peterlechner2008}. 

Bulk intermetallic alloys can be deformed severely using high pressure torsion (HPT). For Zr$_{3}$Al, it was shown that HPT at room temperature leads to a final grain size of approximately 20 nm, but amorphization of a significant volume fraction of the sample has not been encountered \cite{Geist2010}. Furthermore, it was observed that upon HPT deformation, Zr$_{3}$Al exhibits inhomogeneous microstructures, as were also observed in the case of L1$_{2}$ structured Ni$_{3}$Al \cite{Rentenberger2005, Ciuca2009, Ciuca2010}.

Deformation by cold rolling with intermediate foldings (repeated cold rolling - RCR) is an alternative promising deformation route to produce bulk nanocrystalline materials \cite{Wilde2008}. The minimum final grain sizes that can be achieved for a material are often smaller than the corresponding ones after SPD under high pressure (e.g. HPT, equal-channel angular pressing) \cite{Dinda2005}. In addition, RCR can lead to amorphization.

\begin{figure}
\centering
\includegraphics[width=0.45\textwidth]{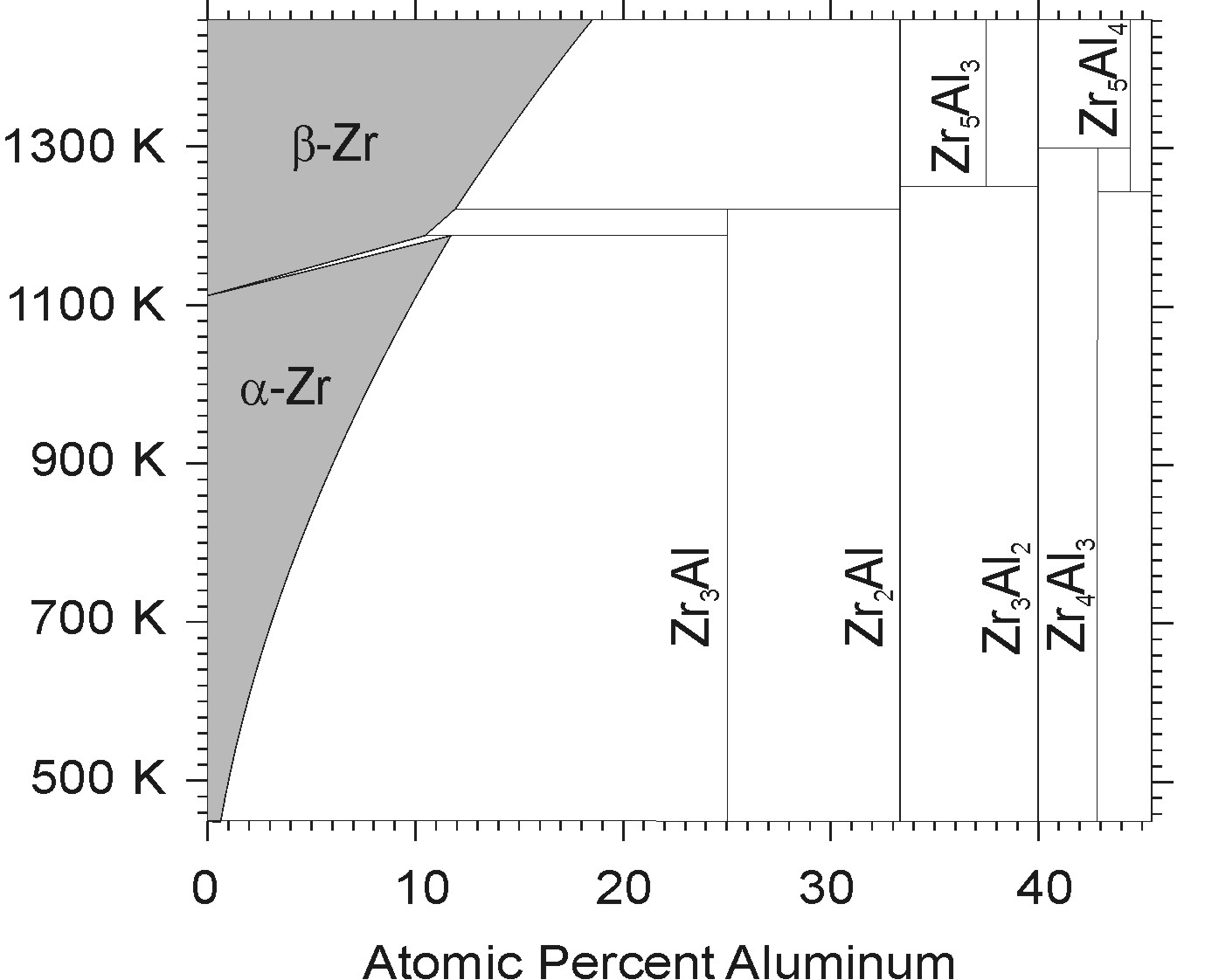}
\caption{The Zr-rich side of the phase diagram of the Zr-Al system. Grey areas are existence regions. The intermetallic phases are all line compounds. The L1$_{2}$-phase at 25at\% Al is stable up to 1261 K, at higher temperatures it decomposes to bcc $\beta$-Zr and Zr$_{2}$Al, which is of the B8$_{2}$-structure \cite{NIMSdb}.}
\label{ZrAl}
\end{figure}

It is the aim of this work to study the effect of RCR on the grain refinement of Zr$_{3}$Al and the behaviour of the refined material upon heating to different temperatures. Crystallization of the amorphous phase is also of interest since the Zr-Al system exhibits several intermetallic phases (cf. Fig. \ref{ZrAl}) and the driving force to form the L1$_{2}$ structure was reported to be rather weak compared to the one of neighbouring phases \cite{Ma1991}.

\section{Experimental Procedure}
\label{exp}

Zr$_{3}$Al was alloyed with an initial composition of Zr with 27 at.\% Al. The alloy was homogenized at 1160 K for 24 h leading to the ordered L1$_{2}$ structure with 10\% residual Zr$_{2}$Al and $\alpha$-Zr. The material was cut to sheets 10mm x 10mm x 0.8mm in size. The thickness of the sheets was reduced to 0.25 mm by repeated cold rolling. Then they were folded and again cold rolled until a thickness of 0.25 mm was achieved. This process was repeated up to 80 times. During rolling the alloy was placed between two spring steel plates.

The deformed material with the highest strain (80 foldings, $\epsilon_{80} = 6 600 \%$) was heated to different temperatures at different heating rates in a differential scanning calorimeter (DSC). For the baseline subtraction, the material was kept at the maximum temperature after heating until the exothermic signal was negligible to avoid exothermic processes in the subsequent run that provided the baseline.

Undeformed samples, samples deformed for 80 foldings and samples heated to 773 K and 973 K at 20 K min$^{-1}$ after deformation for 80 foldings were investigated by X-ray diffraction (XRD) and transmission electron microscopy (TEM). The XRD data were analysed to get information about the coherently scattering domain (CSD) size and the crystal structure of the samples.

The TEM preparation was done by cutting disks suitable for TEM preparation from samples of different states, subsequent grinding and dimpling and finally electropolishing using the same parameters as described in \cite{Geist2008}. Acceleration voltages of 200 kV and 300 kV were used for conventional TEM and high-resolution TEM, respectively.

\section{Experimental Results}
\label{res}

\begin{figure}[t]
\centering
\parbox{0.4\textwidth}{\includegraphics[width=0.4\textwidth]{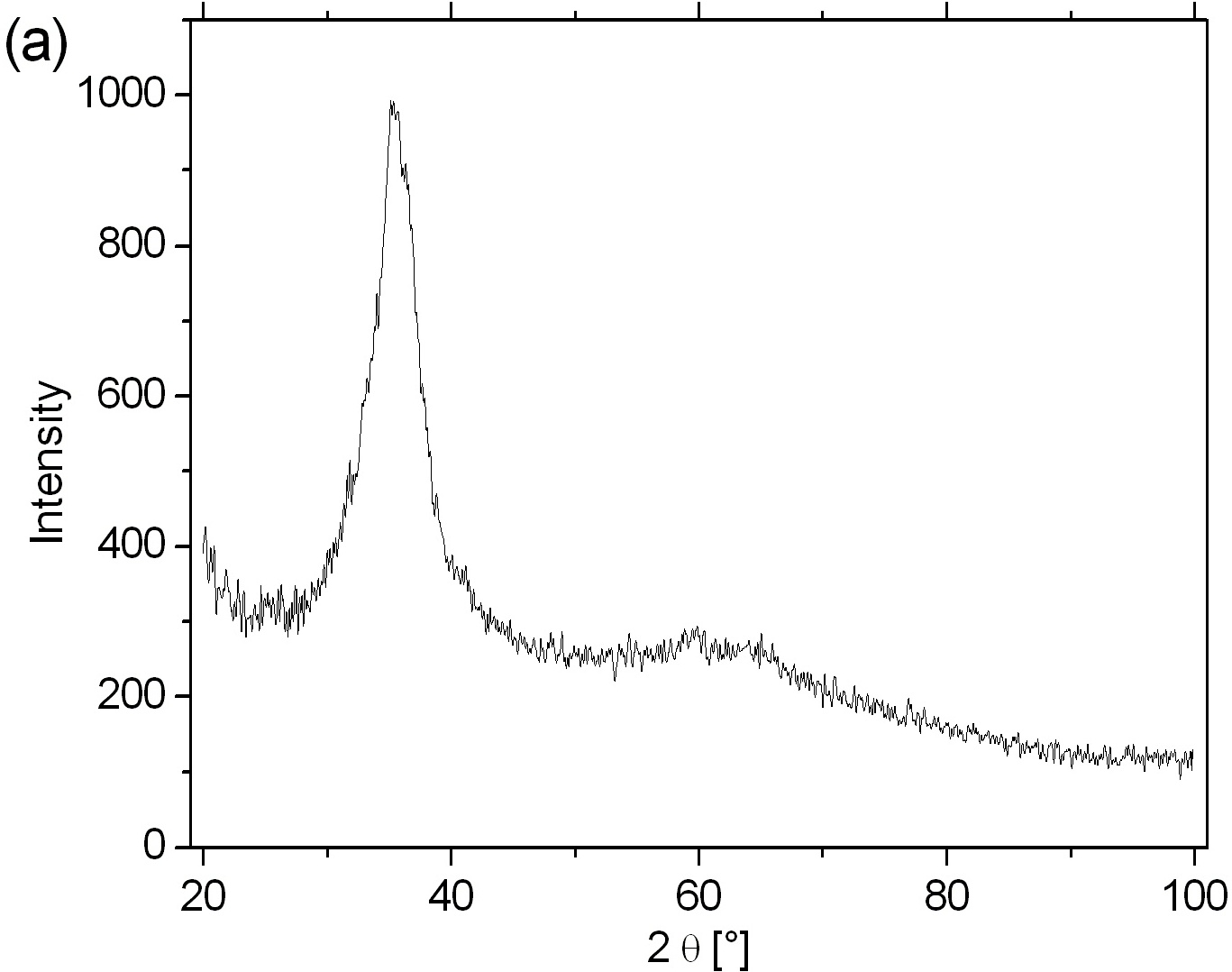}}
\vspace{0.3cm}
\qquad
\begin{minipage}{0.4\textwidth}
\includegraphics[width=\textwidth]{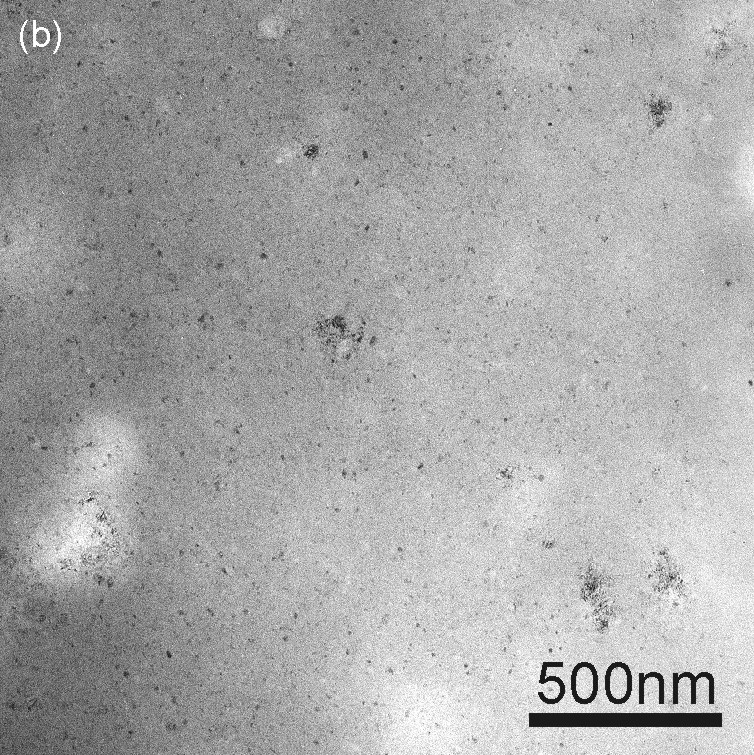}
\end{minipage}
\vspace{0.3cm}
\qquad
\begin{minipage}{0.2\textwidth}
\includegraphics[width=\textwidth]{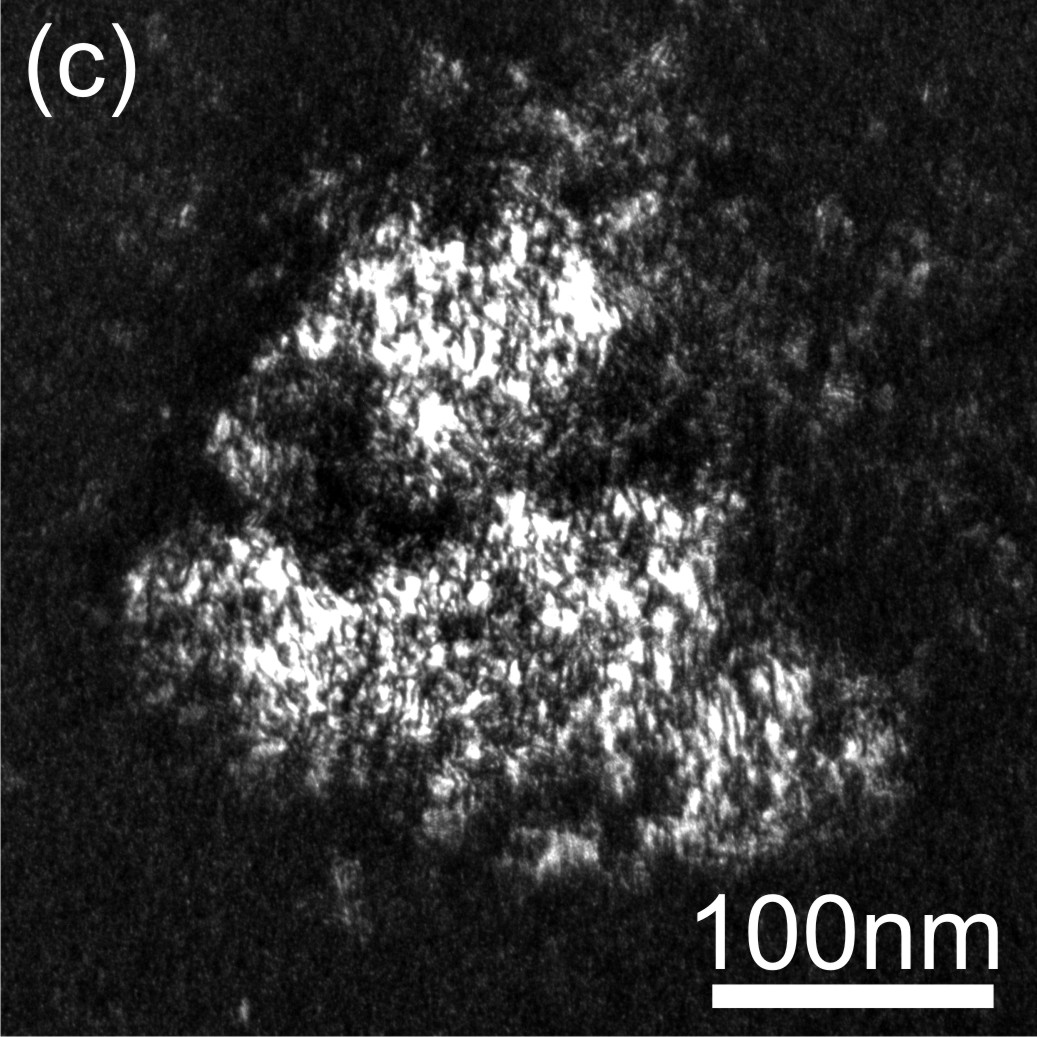}
\end{minipage}
\begin{minipage}{0.2\textwidth}
\includegraphics[width=\textwidth]{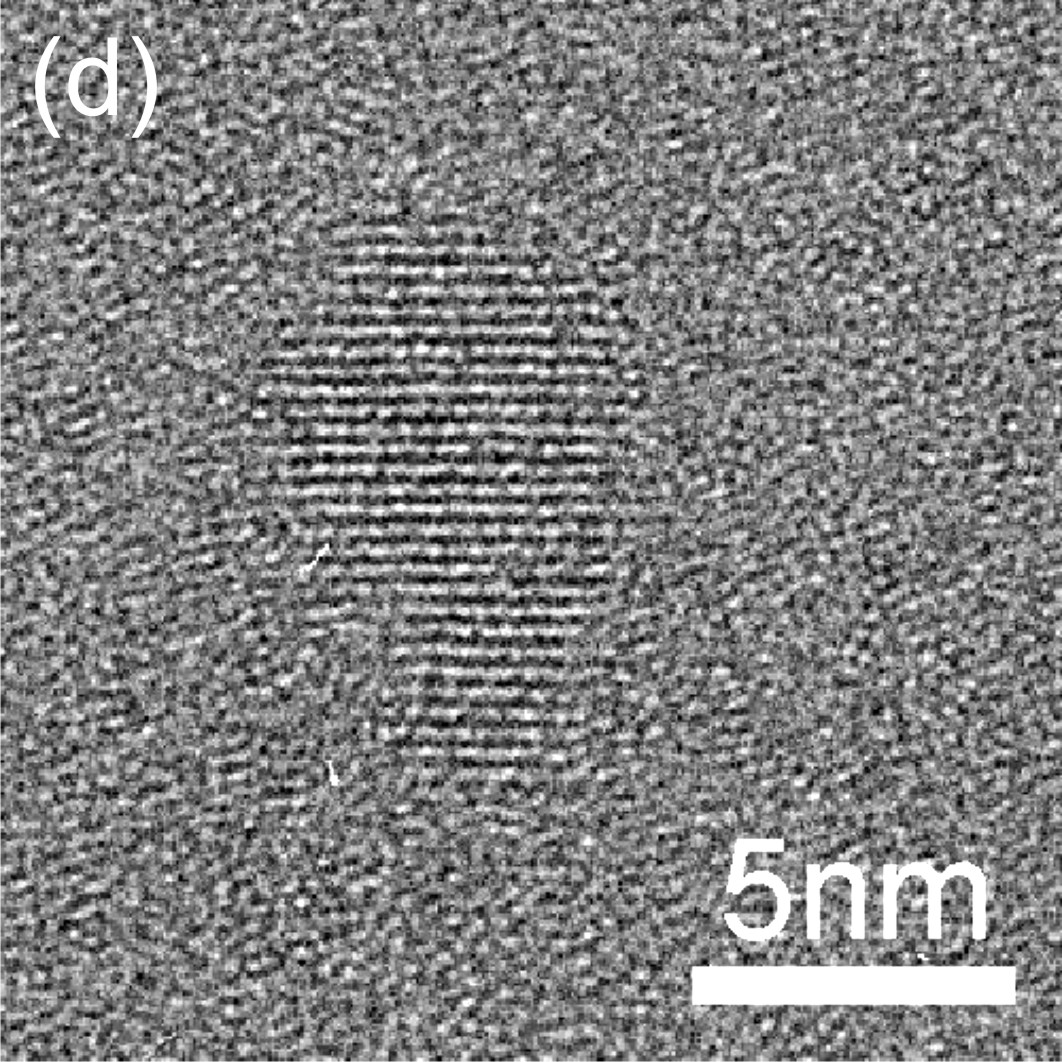}
\end{minipage}
\caption{Structural analysis of the amorphous sample produced by repeated cold rolling with 80 intermediate foldings. (a) XRD spectrum showing two broad peaks indicating a mostly amorphous sample. (b) TEM bright-field image showing large areas of homogeneous intensity. Some larger crystallites (100 - 200 nm) and several small crystallites (black spots, typically $<$10 nm) can be seen in the amorphous matrix. (c) TEM dark-field image of one larger crystallite. The contrast shows that the crystallite contains a lot of structural defects. (d) High-resolution TEM image of a small crystallite embedded in an amorphous matrix.}

\label{asCR}
\end{figure}

TEM analysis of samples deformed for 5, 10 and 20 foldings reveals that already at these relatively low strains ($\epsilon_{5} = 530 \%$, $\epsilon_{10} = 930 \%$, $\epsilon_{20} = 1730 \%$), grain refinement is clearly visible. At 80 foldings ($\epsilon_{80} = 6 600 \%$), XRD shows two broad peaks (cf. Fig. \ref{asCR}a) indicating an amorphous sample. Complementary TEM images taken of the same material yield additional important information (cf. Fig. \ref{asCR}b): a mostly homogeneous intensity distribution with a few remaining crystalline regions that are 100 to 200 nm in diameter (Fig. \ref{asCR}c). From dark-field images, it can be seen that the whole crystalline region is oriented in a similar way, so it is rather one crystallite than an agglomeration of many small crystallites. The complex contrast is an indication for a very high defect density in the crystallite. In addition, small crystallites were identified using high-resolution TEM in the amorphous matrix (Fig. \ref{asCR}d); they show up as dark dots in Fig. \ref{asCR}b (The large ($<$ 500 nm) roundish bright areas originate from thickness variations in the TEM foil developed as an artifact during electropolishing.). Combining the XRD and TEM results, a small crystalline volume fraction of a few percent in a mostly amorphous sample was identified as the sample structure after 80 foldings.

\begin{figure}
\centering
\includegraphics[width=0.5\textwidth]{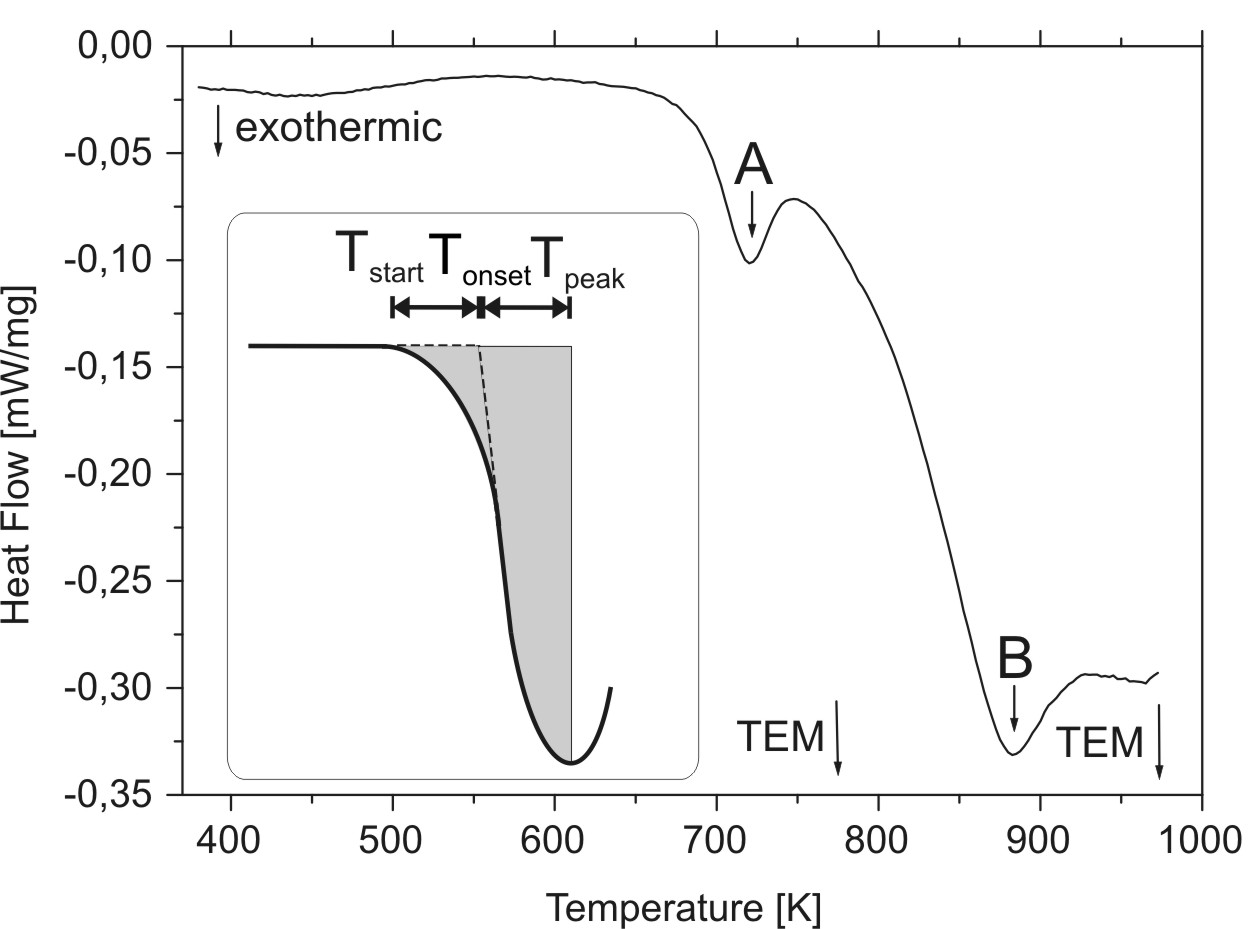}
\caption{DSC curve (heating rate 5 K min$^{-1}$) of Zr$_{3}$Al deformed by 80 foldings. A crystallization peak (marked A) can be seen at 720 K. A large peak (marked B) occurs at higher temperature. The temperatures to which the TEM samples were heated are indicated. A sketch of the integration method is shown in the inset.}
\label{DSC}
\end{figure}

DSC experiments were conducted to clarify the thermal stability of the as-deformed material. Fig.~\ref{DSC} shows a baseline-subtracted DSC curve for a heating rate of 5 K min$^{-1}$. The baseline was obtained by a second run after annealing at 973 K for 30 min. Two exothermic peaks are revealed. The first peak (peak A) that starts at 650 K and has its maximum at 720~K is caused by crystallization of the amorphous phase. To determine the enthalpy of peak A, the following method (cf. inset in Fig. \ref{DSC}) was chosen under the assumption that the peak is symmetric and that peak B has a negligible influence on the low-temperature half of peak A: From the left integration limit $T_{start}=T_{peak}-2(T_{peak}-T_{onset})$ ($T_{onset}$ was determined using the tangent method), the enthalpy between a horizontal baseline and the DSC curve was integrated until $T_{peak}$ (grey area in inset in Fig. \ref{DSC}). This value was taken as the enthalpy of the low-temperature half of the enthalpy of the peak. Due to the above-mentioned assumptions, doubling this value yields the total enthalpy of the peak. The determination of the enthalpies for four different heating rates yielded a crystallization enthalpy of $41\pm$3 J/g. This statistical uncertainty, however, seems to underestimate the real one, that might be as high as 20\%, due to systematic instrumental errors and the assumption of peak B not having an influence on the low-temperature half of peak A. A Kissinger plot using the four different heating rates ranging from 5 K min$^{-1}$ to 50 K min$^{-1}$ leads to an activation enthalpy of $Q=2.60\pm0.10$~eV. The second exothermic peak with a plateau-like behaviour at high temperatures is attributed to a superposition of grain growth and phase transformation processes.

\begin{figure}
\centering
\parbox{0.45\textwidth}{\includegraphics[width=0.45\textwidth]{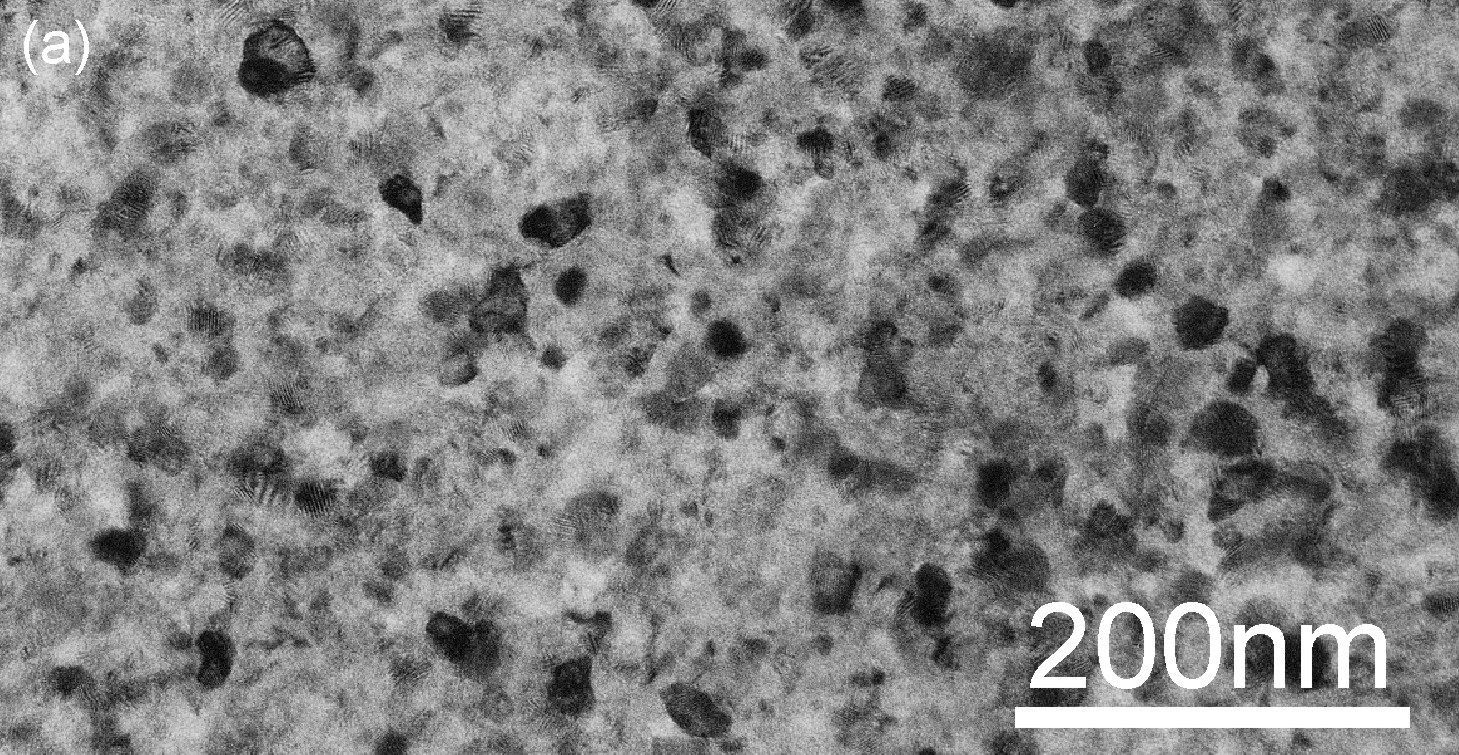}}
\qquad
\begin{minipage}{0.37\textwidth}
\vspace{0.4cm}
\includegraphics[width=\textwidth]{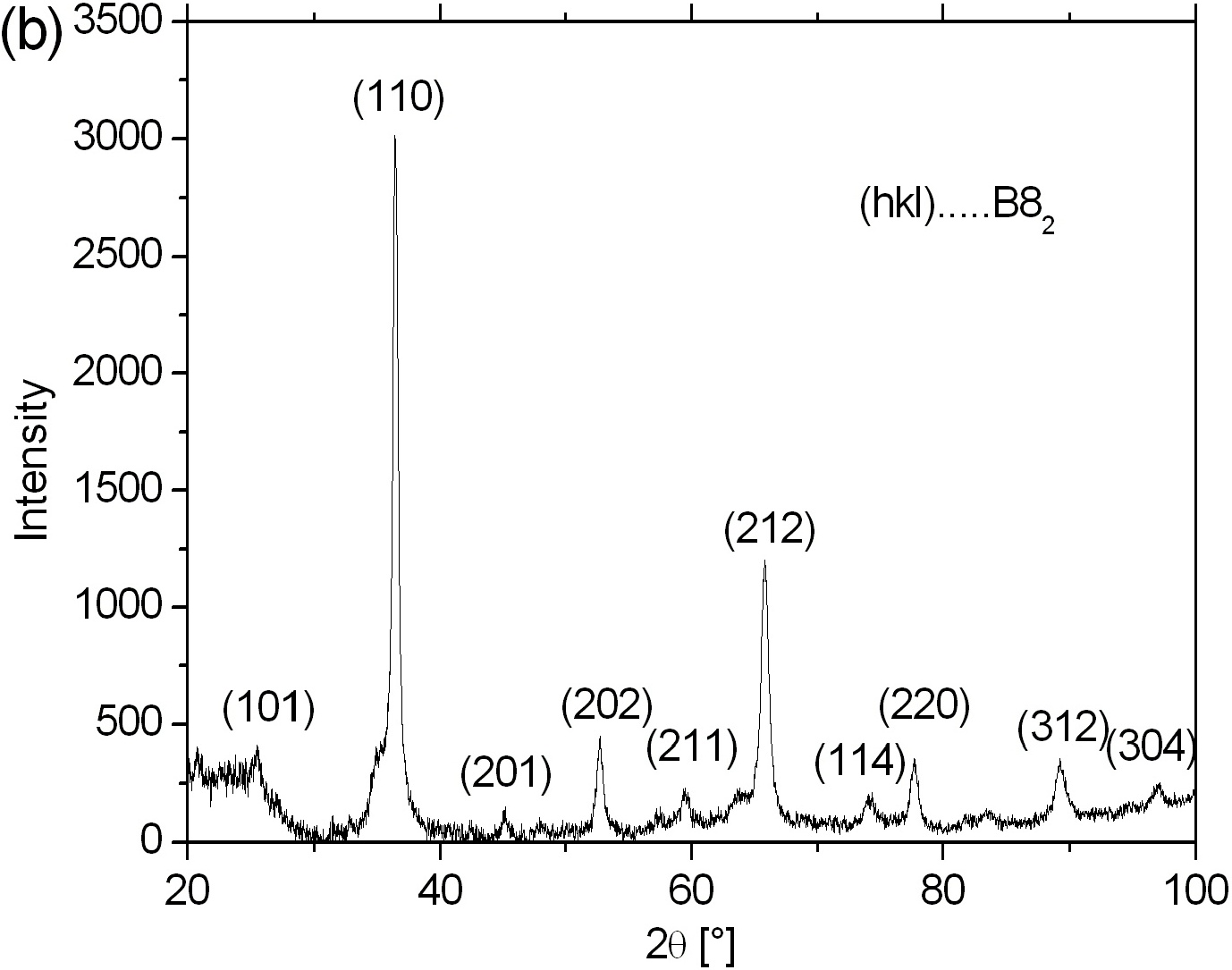}
\end{minipage}
\begin{minipage}{0.45\textwidth}
\vspace{0.2cm}
\includegraphics[width=\textwidth]{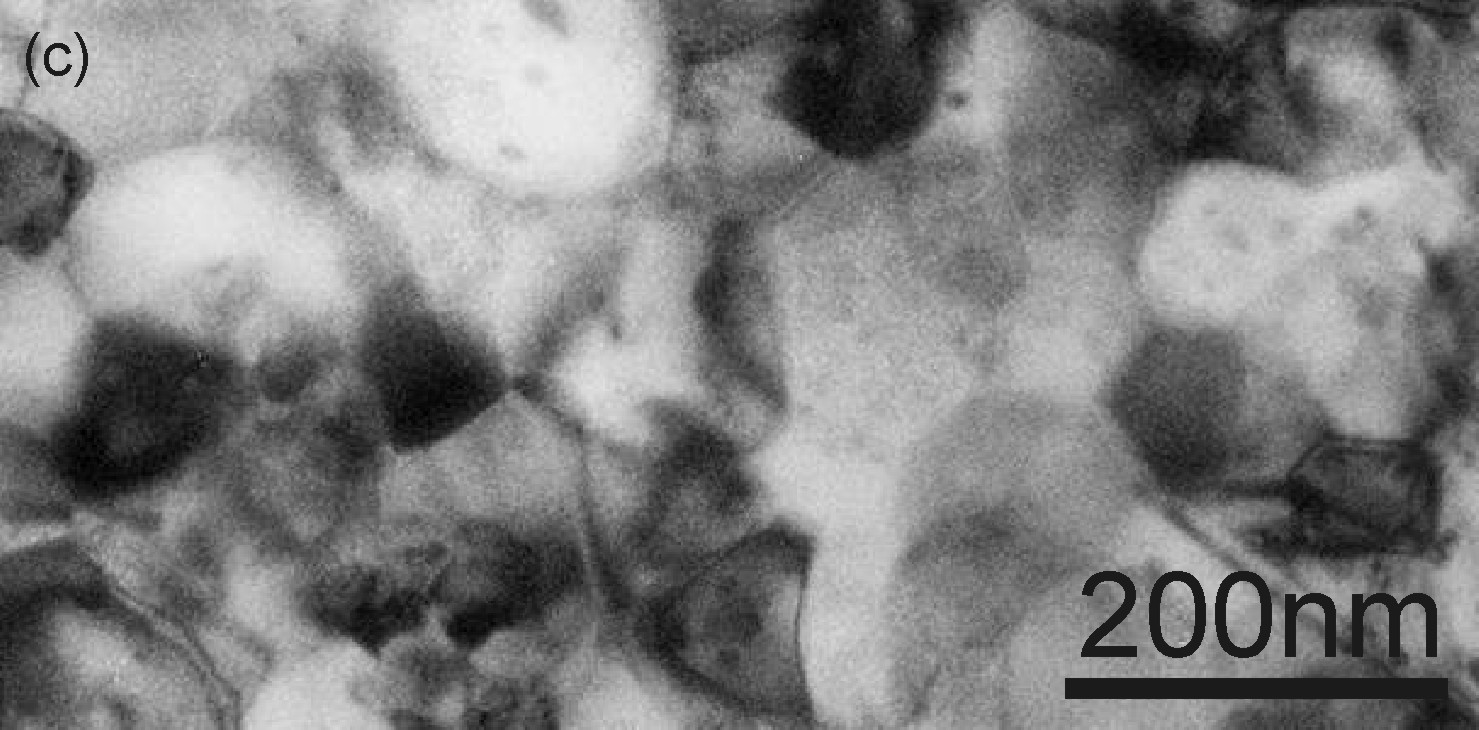}
\end{minipage}
\begin{minipage}{0.37\textwidth}
\vspace{0.3cm}
\includegraphics[width=\textwidth]{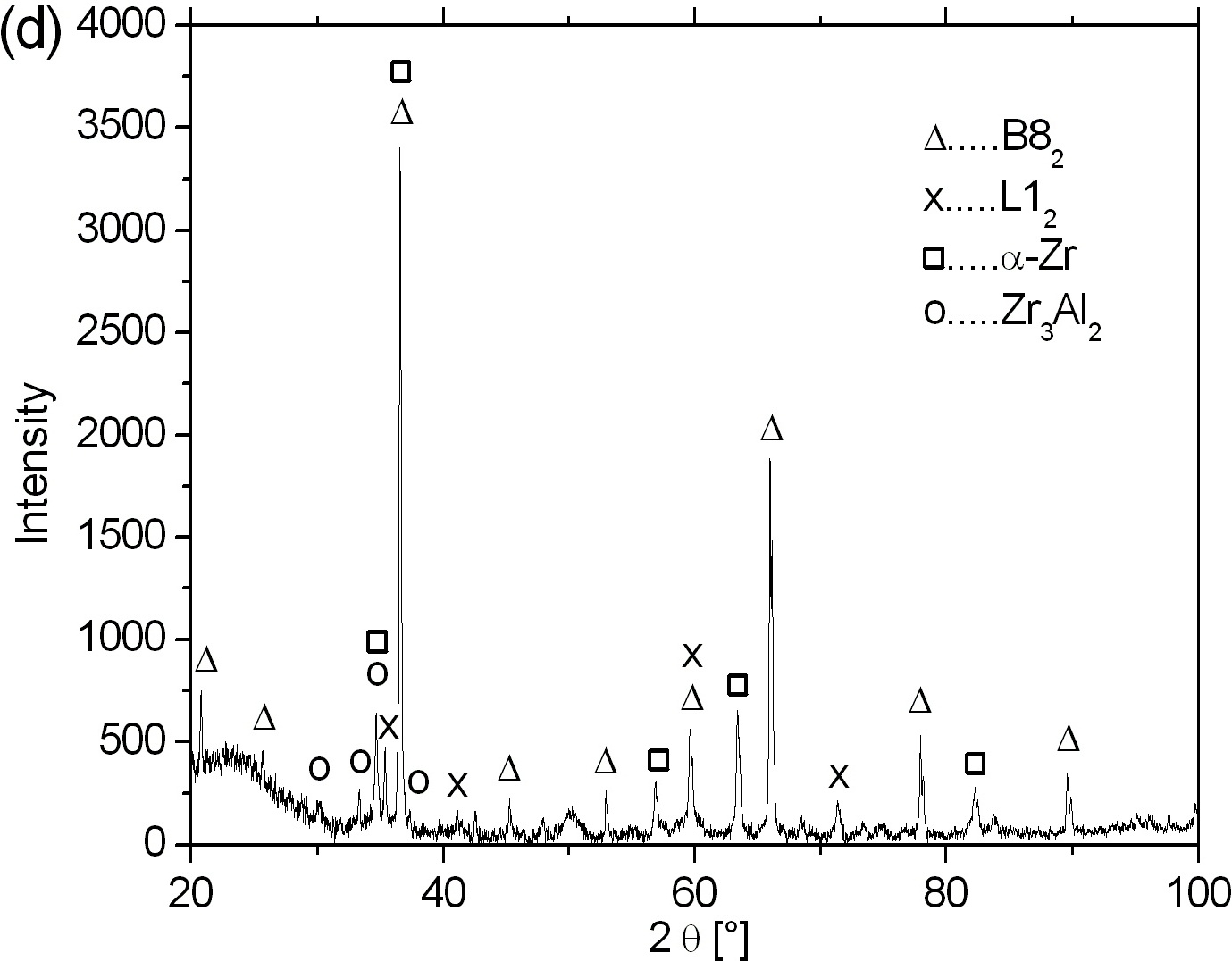}
\end{minipage}
\caption{TEM and XRD of Zr$_{3}$Al deformed by 80 foldings and subsequent heat treatment. For the XRD peaks, the underlying structures and equilibrium compounds are indicated by symbols; the pronounced background at low angles comes from the glass sample holder. (a) TEM image after heating to 773 K. The grain size is 10 - 20 nm. (b) XRD after heating to 773 K. All the peaks correspond to the B8$_{2}$ structure. (c) TEM image after heating to 973 K; defect-free grains with a grain size of $\sim$ 100 nm. (d) XRD after heating to 973 K. The major peaks mainly correspond to the B8$_{2}$ structure; minor peaks to the $\alpha$-Zr, L1$_{2}$ and the equilibrium Zr$_{3}$Al$_{2}$ structure.}
\label{DSC4}
\end{figure}

Using a heating rate of 20 K min$^{-1}$, samples were heated to 773 K and to 973K, i.e. after the crystallization peak and to the maximum temperature of the DSC curve, respectively. Fig. \ref{DSC4}a shows a TEM bright-field image of the sample heated to 773 K revealing nanocrystalline material with a grain size of approximately 10-20 nm.

This grain size estimation is in reasonable agreement with a Williamson-Hall analysis of the CSD size as deduced from the corresponding XRD spectrum (cf. Fig. \ref{DSC4}b) yielding a volume-weighted mean crystalline diameter of 40 nm. Apart from the CSD size, the XRD results show that the B8$_{2}$-structure is clearly predominant in the sample heated to 773 K. All the major peaks can be related to peaks of this structure, which is the equilibrium phase for Zr$_{2}$Al \cite{Massalski1990}. This result was also confirmed by analyzing the corresponding electron diffraction ring patterns using the PASAD software \cite{Gammer2010}. After heating the sample to 973 K, defect-free grains grow to a size of $\sim$ 100 nm as observed in the TEM bright-field image (cf. Fig. \ref{DSC4}c). The corresponding XRD curve (cf. Fig. \ref{DSC4}d) shows that the L1$_{2}$ structure (i.e. the equilibrium structure for Zr$_{3}$Al) starts to form. It should be mentioned that after heating to 973 K at 20 K min$^{-1}$ and immediate subsequent cooling at the same rate, the B8$_{2}$ structure is still predominant.

\section{Discussion}
\label{disc}

TEM analyses of samples deformed by RCR show that pronounced grain refinement takes place at much lower strain levels than it was reported for Zr$_{3}$Al deformed by high pressure torsion \cite{Geist2010}. After 80 foldings ($\epsilon=6 600\%$), at least 90\% of the volume is amorphous and the grain structure is finer than that in the HPT experiment. This is in agreement with the results showing that the grain sizes after RCR can be smaller than those achieved by other SPD techniques \cite{Dinda2005}.

As depicted in Fig. \ref{asCR}b, a homogeneously distributed crystalline debris is retained in the amorphous material. Upon heating to 773 K, the crystallites that are only a few nm large can serve as pre-existing nuclei and lead to the formation of a fine nanocrystalline structure (10 - 20 nm, cf. Fig. \ref{DSC4}a) because of their dense distribution. A similar behaviour was observed for NiTi, that was rendered amorphous by severe plastic deformation and afterwards heated in in-situ TEM experiments \cite{Peterlechner2008}. Peak A of the DSC curve (cf. Fig. \ref{DSC}) is therefore attributed to crystallization and grain growth until impingement. The peak temperature is about 70 K lower than that reported in \cite{Ma1991} for ball-milled amorphous Zr$_{3}$Al. This difference can be explained by the presence of the crystalline debris in our samples. A similar behaviour was observed in NiTi with a peak temperature reduction of 100 K \cite{Peterlechner2008}. Therefore, it is concluded that the higher peak temperature reported for Zr$_{3}$Al in \cite{Ma1991} indicates that the material did not contain a crystalline debris. Peak B of the DSC curve (cf. Fig. \ref{DSC}) is interpreted to be caused by grain growth and phase transformations between different intermetallic phases (cf. Fig. \ref{DSC4}c and d).

By experimental and theoretical evaluation, Ma et al \cite{Ma1993} deduced an enthalpy release caused by devitrification of Zr$_{3}$Al of $\sim$ 85 J/g, which is about twice the amount that is determined here. Possible reasons for the higher crystallization enthalpy are an additional contribution of grain growth, which is part of DSC peak B (cf. Fig. \ref{DSC}) in this work or a principal difference in the structure of the amorphous samples caused by the different processing conditions (strains, strain rates, deformation temperatures,...)

X-ray results show that the equilibrium L1$_{2}$ structure does not form in the initial crystallization process since only B8$_{2}$ structured Zr$_{2}$Al was detected after heating to 773 K (cf. Fig. \ref{DSC4}b). It is well established that nanocrystalline material can have a different crystal structure than its coarse crystalline counterpart at the same composition and temperature \cite{Mayo2003, Jiang2008}. In the present case, antisites or vacancies on Al sites of the ordered B8$_{2}$ structure could in addition compensate for the difference between the nominal as-cast composition and that of stoichiometric Zr$_{2}$Al. From the DSC curve, the formation enthalpy of the present B8$_{2}$ structure (containing a high number of point defects causing excess enthalpy) is determined to lie between the one of the amorphous phase and the one of the L1$_{2}$ structure as given in \cite{Ma1993}. This makes the transformation path $amorphous \rightarrow B8_{2} \rightarrow L1_{2}$ thermodynamically reasonable. A comparison with data of Zr$_{5}$Al$_{3}$ and Zr$_{5}$Al$_{4}$ favors the substitution with vacancies \cite{Nandedkar1982}. Assuming a vacancy density that compensates for the aluminium deficiency in the formation of the B8$_{2}$ structure and ignoring grain size effects causing excess enthalpy gives a reasonable upper limit of 2.7 eV for the formation enthalpy of a vacancy.

After heating to 973 K, the Zr$_{2}$Al phase is still predominant, but some $\alpha$-Zr, L1$_{2}$ structured Zr$_{3}$Al and a little Zr$_{3}$Al$_{2}$ of the space group P4$_{2}$/mnm starts to form (cf. Fig. \ref{DSC4}d).

The heat treatments presented in this work resulted in a very fine nanocrystalline structure dominated by the B8$_{2}$ phase (10 - 20 nm). It is suggested that heat treatment of amorphous samples at low temperatures ($\sim$ 770 K) for several hours or even days might give rise to the formation of the L1$_{2}$ structure without causing excessive grain growth. It is concluded that the crystalline debris resulting from the amorphization by RCR is important for reducing the grain size of the crystallized material by providing a dense network of homogeneously distributed pre-existing nuclei.

\section{Conclusions}

\begin{itemize}
\item{Zr$_{3}$Al can be rendered amorphous by repeated cold rolling as concluded from the X-ray results. The TEM studies show that the deformed material contains a crystalline debris that is not resolved by X-ray diffraction.}
\item{Heating to 773 K leads to the crystallization of small (10 - 20 nm in diameter) B8$_{2}$-structured nanocrystals. In this nanocrystalline structure, the non-equilibrium Zr$_{2}$Al phase is clearly predominant.}
\item{It is concluded that during heating, the crystalline debris (of the Zr$_{3}$Al phase retained in the amorphous phase) acts as nuclei for the crystallization process of the non-equilibrium phase Zr$_{2}$Al.}
\item{When heating to 973 K, the equilibrium L1$_{2}$ phase starts to form and the crystals grow to a size of about 100 nm.}
\end{itemize}

\section*{Acknowledgments}
\label{ackn}
The authors thank Prof. Erland M. Schulson (Dartmouth College, New Hampshire, USA) for the kind provision of Zr$_3$Al. D.G., H.P.K. and C.R. acknowledge support by the research project `Bulk Nanostructured Materials' within the research focus `Materials Science' of the University of Vienna and by the Austrian Science Fund (FWF):[P22440]. D.G. acknowledges the support by the `NIMS internship program' of the National Institute for Materials Science, Tsukuba, Japan and by the IG `Experimental Materials Science - Nanostructured Materials', a  college for PhD students at the University of Vienna.

\appendix

\bibliographystyle{unsrt}

\end{document}